\documentclass[pdftex,twocolumn,epjc3]{svjour3}          

\RequirePackage[T1]{fontenc}

\smartqed  

\RequirePackage{graphicx}
\RequirePackage{mathptmx}      
\RequirePackage{flushend}
\RequirePackage[numbers,sort&compress]{natbib}
\RequirePackage[colorlinks,citecolor=blue,urlcolor=blue,linkcolor=blue]{hyperref}
\usepackage{tablefootnote}
\usepackage{hyphenat}

\journalname{Eur. Phys. J. C}

\usepackage[normalem]{ulem} 

\def\Re{$^{187}$Re}
\def\Ho{$^{163}$Ho}

\def\agre{AgReO$_4$}

\def\mum{$\mu$m}

\def\mus{$\mu$s}

\def\ero{Er$_2$O$_3$}

\def\Ho{$^{163}$Ho}

\def\Er{$^{162}$Er}

\begin{document}

\title{Impact of embedded \Ho\ on the performance of the transition-edge sensor microcalorimeters of the HOLMES experiment}

\institute{
National Institute of Standards and Technology (NIST), Boulder, Colorado, USA\label{addr1}
\and
University of Colorado, Boulder, Colorado, USA\label{addr2}
\and
Istituto Nazionale di Fisica Nucleare (INFN), Sezione di Genova, Genova, Italy\label{addr3}
\and
Dipartimento di Fisica, Universit\`a di Genova, Genova, Italy\label{addr4}
\and
Dipartimento di Fisica, Universit\`a di Milano-Bicocca, Milano, Italy\label{addr5}
\and
Istituto Nazionale di Fisica Nucleare (INFN), Sezione di Milano Bicocca, Milano, Italy\label{addr6}}

\date{Received: date / Accepted: date}

\author{Douglas Bennett\thanksref{addr1}
  \and
  Matteo Borghesi\thanksref{addr5,addr6}
  \and
  Pietro Campana\thanksref{addr5,addr6}
  \and
  Rodolfo Carobene\thanksref{addr5,addr6}
  \and
  Giancarlo Ceruti\thanksref{addr6}
  \and
  Matteo De Gerone\thanksref{addr3}
  \and
  Marco Faverzani\thanksref{addr5,addr6}
  \and
  Lorenzo Ferrari Barusso\thanksref{addr3,addr4}
  \and
  Elena Ferri\thanksref{addr6}
  \and
  Joseph Fowler\thanksref{addr1}
  \and
  Sara Gamba\thanksref{addr5,addr6}
  \and
  Flavio Gatti\thanksref{addr3,addr4}
  \and
  Andrea Giachero\thanksref{addr5,addr6}
  \and
  Marco Gobbo\thanksref{addr5,addr6}
  \and
  Danilo Labranca\thanksref{addr5,addr6}
  \and
  Roberto Moretti\thanksref{addr5,addr6}
  \and
  Angelo Nucciotti\thanksref{e1,addr5,addr6}
  \and
  Luca Origo\thanksref{addr6}
  \and
  Stefano Ragazzi\thanksref{addr5,addr6}
  \and
  Dan Schmidt\thanksref{e2,addr1}
  \and
  Daniel Swetz\thanksref{addr1}
  \and
  Joel Ullom\thanksref{addr1,addr2}
}
\thankstext{e1}{e-mail: angelo.nucciotti@mib.infn.it}
\thankstext{e2}{e-mail: dan.schmidt@nist.gov}

\maketitle

\begin{abstract}
We present a detailed investigation of the performance of transition-edge sensor (TES) microcalorimeters with $^{163}$Ho atoms embedded by ion implantation, as part of the HOLMES experiment aimed at neutrino mass determination. The inclusion of $^{163}$Ho atoms introduces an excess heat capacity due to a pronounced Schottky anomaly, which can affect the detector's energy resolution, signal height, and response time. We fabricated TES arrays with varying levels of $^{163}$Ho activity and characterized their performance in terms of energy resolution, decay time constants, and heat capacity. The intrinsic energy resolution was found to degrade with increasing $^{163}$Ho activity, consistent with the expected scaling of heat capacity. From the analysis, we determined the specific heat capacity of $^{163}$Ho to be $(2.9 \pm 0.4 \mathrm{(stat)} \pm 0.7 \mathrm{(sys)})$ J/K/mol at $(94 \pm 1)$\,mK, close to the literature values for metallic holmium. No additional long decay time constants correlated with $^{163}$Ho activity were observed, indicating that the excess heat capacity does not introduce weakly coupled thermodynamic systems. These results suggest that our present TES microcalorimeters can tolerate $^{163}$Ho activities up to approximately 5 Bq, with only about a factor of three degradation in performance compared to detectors without $^{163}$Ho. For higher activities, reducing the TES transition temperature is necessary to maintain or improve the energy resolution. These findings provide critical insights for optimizing TES microcalorimeters for future neutrino mass experiments and other applications requiring embedded radioactive sources. The study also highlights the robustness of TES technology in handling limited amounts of implanted radionuclides while maintaining high-resolution performance.

\end{abstract}

\section{Introduction}
Calorimetric measurement of the fraction of energy released in nuclear beta decays that is not carried away by neutrinos was originally proposed to overcome the limitations of ex\-ternal-source spectrometric neutrino mass experiments with tritium \cite{simpson_measurement_1981} and \Ho\ \cite{derujula_calorimetric_1982}.
Pioneering neutrino mass calorimetric experiments have been performed with tritium ions implanted in Si(Li) \cite{simpson_measurement_1981} and HPGe \cite{hime_evidence_1989} detectors, with \Ho\ ions implanted in HPGe \cite{laegsgaard_capture_1984} detectors, and with a high\hyp{}temperature gas proportional detector containing a gaseous organometallic holmium compound \cite{hartmann_high_1992}.
In those early experiments based on standard detector technologies, sensitivity was limited by the energy resolution and possibly by solid-state effects \cite{franklin_appearance_1995}. 
 The advent of sensitive low-temperature detectors (LTDs) has provided new opportunities to improve
calorimetric experiments \cite{nucciotti_use_2016b} and has expanded the choice of source isotopes.
Besides some attempts to use  ion-implanted tritium  \cite{erhardt_transition-edge_2000,lowry_beta_1993}, early neutrino mass experiments with LTDs focused on the low $Q$ beta decay of \Re\ in both metallic rhenium \cite{fontanelli_data_1999} and dielectric \agre\ \cite{sisti_new_2004} energy absorbers. 
However, the very low specific activity of rhenium (\Re, half-life $\sim4\times10^{10}$\,years) made scaling up these experiments impractical~\cite{nucciotti_use_2016b}. 
In contrast, the much short\-er half\hyp{}life of electron\hyp{}capture\hyp{}decaying \Ho\ (about 4570\,years \cite{baisden_measurement_1983}) makes it preferable for LTD-based calorimetric experiments, requiring only $2\times10^{11}$ nuclei per 1\,decay/sec.
Several approaches have been explored for the inclusion of the \Ho\ source in the absorbers, including the encapsulation by epoxy drops contained in folded tin foils \cite{gatti_calorimetric_1997}, the inclusion of holmium in superconducting yttrium compounds \cite{gastaldo_superconducting_2004}, or the drying of a holmium\hyp{}containing solution dripped onto nanoporous gold \cite{croce_development_2016}.
However, the most promising technique to date remains ion implantation of \Ho, adopted by the ECHo \cite{gastaldo_electron_2017} and HOLMES \cite{alpert_holmes_2015b} neutrino mass experiments.
Radioactive sources have been embedded in LTD absorbers -- mainly via ion implantation -- for applications such as $^7$Be electron-capture studies for sterile neutrino searches \cite{friedrich_limits_2021}, astrophysical process investigations \cite{fretwell_direct_2020}, 
and total decay energy spectroscopy for metrology, safeguards, and other purposes \cite{koehler_low_2021}.
Recently, tritium generated in LiF LTDs by neutron capture has also been used for sterile neutrino searches \cite{lee_lab-scale_2022}.

\section{Transition-edge Sensor Microcalorimeters for HOLMES}
The HOLMES experiment uses transition-edge sensor (TES) microcalorimeters \cite{irwin_transition-edge_2005}
whose design and fabrication details are described here and in \ref{app:fab}. 
In this paper we also investigate how the inclusion of \Ho\ atoms affects the detector response.

\begin{figure}[t!]
 \begin{center}
 \includegraphics[width=0.45\textwidth]{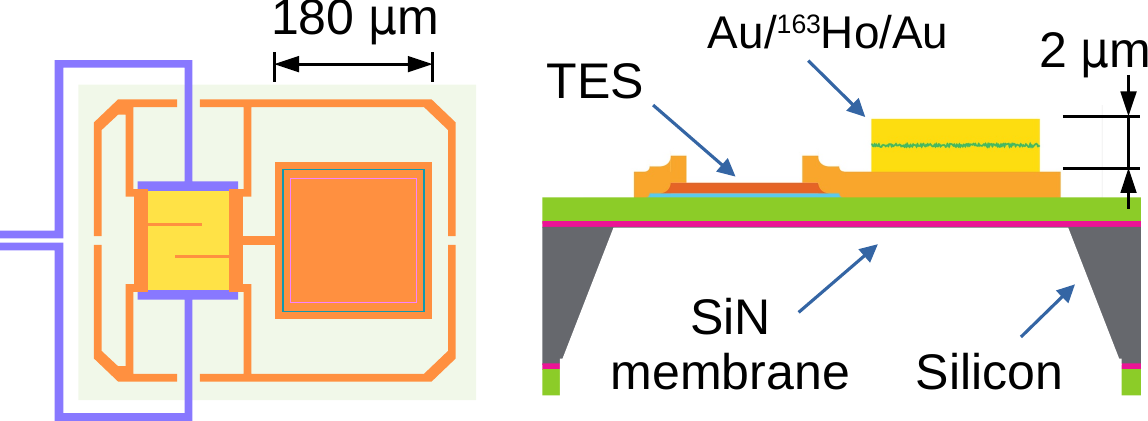} 
 \end{center}
 \caption{
  Left: Layout of the TES microcalorimeter, showing the side-by-side Mo/Cu bilayer sensor and gold absorber, with a copper perimeter for enhanced thermal conduction\cite{hays-wehle_thermal_2016}. Right: Schematic (not to scale) of the micromachined TES microcalorimeter with embedded \Ho\ as used in this work.}
 \label{fig:tes}
 \label{fig:fab}
\end{figure}

The TES \cite{irwin_transition-edge_2005} microcalorimeters used for this work are specifically designed \cite{alpert_high-resolution_2019b} to meet the experimental requirements of HOLMES. Mo/Cu bilayer TESs are tuned to have critical temperatures around $95\,$mK and are thermally coupled via a thin-film copper link to a 180$\times$180$\times 2\,\mu$m$^3$ gold absorber, in which the $^{163}$Ho is embedded by ion implantation (Fig.\,\ref{fig:tes}). 
HOLMES microcalorimeters are designed with the absorber positioned beside the sensor to prevent proximization \cite{faverzani2016holmes}.
Detectors are suspended on a SiN$_\mathrm{x}$ membrane to provide a finite thermal link $G$ to the thermal bath, which is kept at a constant 
temperature of $40\,$mK.
Additional copper banks are inserted to increase $G$ and to control the TES resistance and excess noise \cite{hays-wehle_thermal_2016,ullom_characterization_2004,alpert_high-resolution_2019b}.

The detectors operate as thermal equilibrium calorimeters \cite{mccammon_thermal_2005}, where all energy deposited in the absorber is converted into a thermal signal. In the simplest model, the system is treated as a single heat capacity -- the sum of absorber and TES electron contributions -- characterized by the TES electron temperature. Both signal amplitude and decay time constant depend on this total heat capacity. For accurate performance predictions, electrothermal feedback from constant-voltage TES biasing must be included, as described by the full electrothermal model \cite{irwin_transition-edge_2005}. Here, we use its small-signal approximation.

Future Ho-163-based neutrino mass experiments aiming at sub-eV sensitivities will need to accumulate statistics exceeding $10^{13}$ decays. To minimize the number of detectors and the total measurement time, it would be desirable to achieve per-detector activities of several tens or even hundreds of decays per second \cite{nucciotti_statistical_2014b}. Therefore, it is critical to assess the impact of doping on detector performance.
In fact, sensitivity would be adversely affected by excessive deterioration of energy resolution or increase in dead time \cite{nucciotti_statistical_2014b,borghesi_matrix_2022b}.
The experiment's dead time is determined by the signal fall time, as any new event occurring during the decay of a previous pulse results in both events being discarded from analysis. Since the fall time is proportional to the heat capacity ($C$), an increased heat capacity leads to longer dead times. Furthermore, the heat capacity directly impacts the achievable energy resolution: for negligible readout noise and in the regime of strong electrothermal feedback, the energy resolution scales as $\sqrt{C}$, so a larger heat capacity results in poorer energy resolution \cite{irwin_transition-edge_2005}. 
In contrast, the rise time of the signals is primarily determined by the electrical cutoff set by $L/R_0 \approx 20\,\mu$s, where $R_0$ is the sensor resistance at its operating point and $L$ is the inductance of the bias circuit. 
The rise time is related to the detector resolving time and therefore determines the level of background at the end-point caused by \Ho\ decays occurring too close in time to be resolved \cite{nucciotti_statistical_2014b}.

Holmium it is known to exhibit an excess heat capacity due to a pronounced Schottky anomaly \cite{krusius_calorimetric_1969}.
Hyperfine \cite{krusius_calorimetric_1969} and crystalline field \cite{williams_crystal-field_1969} splittings are caused by $4f$ atomic electrons interacting with \Ho\ nuclear spins ($I=7/2$) and with electric field gradients in the $fcc$ lattice of the gold host, respectively.
These splittings induce Schottky anomalies in the heat capacity of the \Ho\ nuclei which are expected to peak between $0.1$\,K and 1\,K, i.e., near the operating temperature of HOLMES detectors.
As a result, microcalorimeters loaded with large amounts of $^{163}$Ho are expected to show degraded performance.
By analyzing detector parameters -- specifically energy resolution and signal decay time -- at varying $^{163}$Ho activity, we can infer the contribution of holmium to the total heat capacity of the absorber.

Depending on the strength of the coupling between the detector electronic system and the \Ho\ nuclei ensemble, this excess heat capacity can cause either an increase of the total detector heat capacity or the appearance of an additional weakly coupled thermodynamic system. 
Therefore, depending on the \Ho\ concentration and on the detector operating temperature, the \Ho\ doping can manifest as a reduction in signal height, a worsening of detector resolution, a slowing of detector response or a complex signal shape with additional long decay times.

The heat capacity of metallic holmium has been measured to be  3.8\,J/K/mol \cite{krusius_calorimetric_1969} at 95\,mK.
The amount of $^{163}$Ho to be added in the detectors for a unit activity is $2.1\times 10^{11}$\,nuclei, or $3.5\times10^{-13}$\,moles, corresponding to an additional heat capacity of about $1.3\times10^{-12}$\,J/K.
For comparison, the total heat capacity of HOLMES microcalorimeters at the same temperature is estimated to be about $8\times10^{-13}$\,J/K \cite{alpert_high-resolution_2019b}. This means that the heat capacity added by 1\,Bq of \Ho\ is already larger than the intrinsic heat capacity of the detector. It should be noted that using the heat capacity of metallic holmium for holmium implanted in a gold matrix represents an extreme case; finding the actual contribution is the purpose of this work.

A first investigation of the impact of \Ho\ doping in magnetic metallic microcalorimeters was reported by the ECHo collaboration \cite{gastaldo_characterization_2013,herbst_specific_2021}.

The  HOLMES detectors are manufactured using a combination of thin-film fabrication techniques and silicon micromachining.  The \Ho\ atoms are embedded in the detectors during fabrication using ion implantation and then encapsulated with a final layer of Au.  Detectors are grouped in 64 pixel arrays contained on individual silicon die (Fig.\,\ref{fig:array}).  Details of the fabrication and implantation are given in \ref{app:implant} and \ref{app:fab}.

\section{Experimental Methods and Detector Characterization}
\label{sec:exp_meth}
The data presented in this work are taken with microcalorimeters belonging to three arrays. 
The first array (Array 0) was fabricated without ion implantation to verify that the fabrication process developed for encapsulation yields devices with performance matching those in \cite{alpert_high-resolution_2019b}, which were produced in a single step using DRIE micromachining at NIST.
The other two arrays -- Arrays 1 and 2 -- differ in their ion implantation protocols (see \ref{app:implant}), with detector activities ranging from 0 up to about 1\,Bq and 0.6\,Bq, respectively.

\begin{figure}[!t]
 \begin{center} 
 \includegraphics[width=0.45\textwidth]{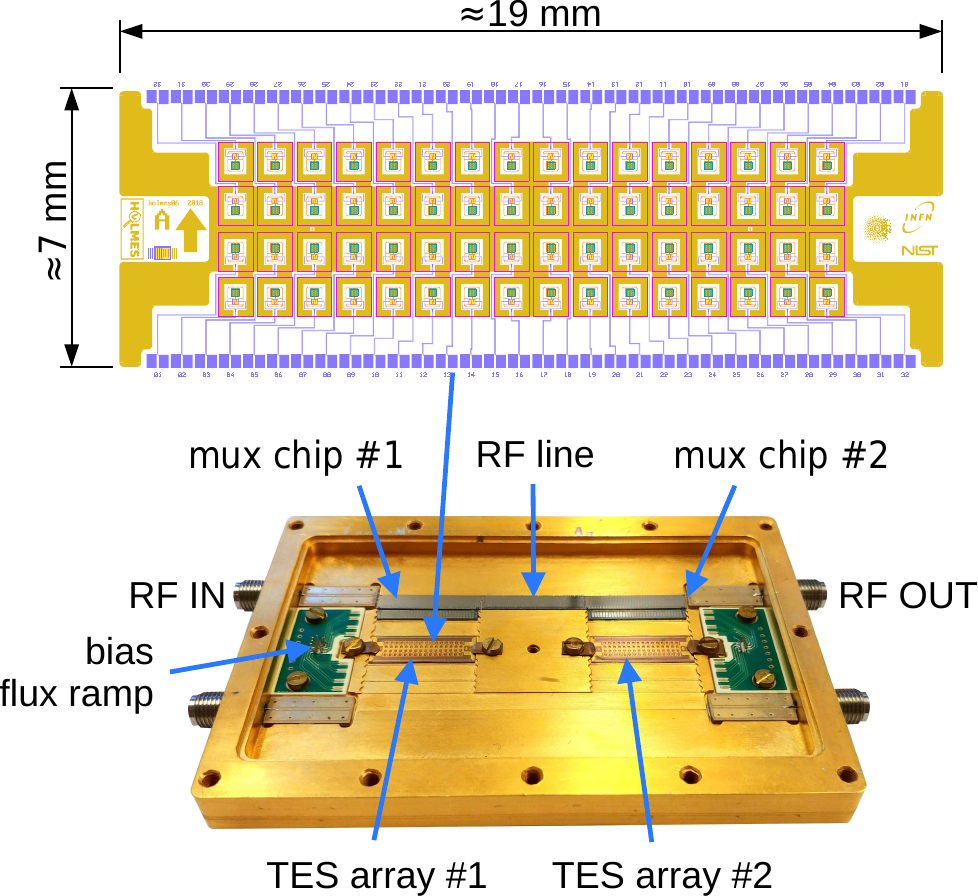}
\end{center}
\caption{Top: Layout of one of the two $16\times4$ pixel arrays fabricated on $14\times19$\,mm$^2$ chips. Each cell reproduces the layout shown in Fig.\,\ref{fig:fab} (Left). Bottom: Copper box used for characterizing Arrays 0 and 1. The TES arrays are mounted at the center. Only the upper half of the arrays (32 microcalorimeters) is instrumented with the bias network and microwave multiplexer chips. The multiplexers have their feedlines aligned with the SMA connectors and the central interconnecting feedline. The box is closed with a light-tight cover featuring two openings above the array, each shielded by a 6\,\mum\ aluminum foil, to allow irradiation with an external X-ray source. For Array 2, a new box accommodating a single array was used (see figure in \cite{alpert_most_2025}).}
\label{fig:holders}
 \label{fig:array}
\end{figure}

All measurements were conducted in a dilution refrigerator, where the copper boxes hosting the arrays (see Fig.\,\ref{fig:holders}) were maintained at a stable temperature of 40 mK. 
For detector calibration, a fluorescence source was used, consisting of a primary $^{55}$Fe source that irradiated a target composed of sodium chloride, calcium carbonate, and aluminum. The resulting K$\alpha$ X-rays from aluminum (1486\,eV), chlorine (2622\,eV), calcium (3690\,eV), and manganese (5900\,eV) were used as calibration points.
For Arrays 1 and 2 containing \Ho, the spectra also exhibit prominent peaks corresponding to the calorimetric detection of atomic de-excitation energy from dysprosium atoms left with core holes following electron capture \cite{alpert_most_2025}. The most intense are the N1 and M1 capture peaks, at approximately 412\,eV and 2041\,eV, respectively.
Each detector is characterized by acquiring IV curves at various heat bath temperatures $T_{b}$, measured using a calibrated RuO$_x$ thermometer. From these measurements, the thermal conductance $G(T)$ is extracted and parametrized as $G(T) = n g T^{n-1}$, where $n$ and $g$ are obtained from a fit to the data.
The optimal operating point parameters, specifically the TES current $I_0$ at a resistance $R_0$ set to approximately 30\% of the normal-state value, are also determined. 
For each detector, the operating temperature $T_0$ is then obtained from $P_{J0} = I_0^2 R_0$ using the relation $P_{J0} = g (T^n-T^n_{b})$ and $T_0 \approx 95$\,mK.

 \begin{table*}
  \centering
   \caption{Summary of array parameters and performance. Time constants $\tau_{rise}$ and $\tau_{dec}$ refer to low-energy signals (e.g., Al X-rays or N1 \Ho\ capture). Except for $G$, all values are for TESs with minimal or no implanted activity. Intrinsic detector resolutions $\Delta E_\mathrm{0}$ are calculated from Eq.\,(\ref{eq:nep}), while $\Delta E_\mathrm{FWHM}$ are the FWHMs obtained by fitting the Mn K$\alpha$ peaks. Note that performance comparisons should account for differences in bath and operating temperatures.}
  \label{tab:results}
 \begin{tabular*}{\textwidth}{@{\extracolsep{\fill}}cccccc@{}}
  \hline
    Array & $\Delta E_\mathrm{0}$ [eV] & $\Delta E_\mathrm{FWHM}$ [eV] & $\tau_{rise}$ [\mus] & $\tau_{dec}$ [\mus] & $G$ [pW/K]$^a$\\
    \hline
  TESs in \cite{alpert_high-resolution_2019b} & 3.3 & $4.5\pm0.1$ & 13 & $54(220)$$^b$ & 600\\
  0$^c$ & 3.7 & $4.2\pm0.1$ & 15 & 300 & $387\pm57$\\
  1$^d$ & 4.8 & $5.8\pm0.1$ & 9 & 360 & - - - \\
  2$^e$ & 5.0 & 6.5$\pm$0.1 & 8 & 600 & $171\pm26$\\
  \hline
  \end{tabular*}
  \begin{flushleft}
  \footnotesize
  $^a$ See also comments in \ref{app:fab}\\ 
  $^b$ Fast pulses in \cite{alpert_high-resolution_2019b} are modeled with two exponential decays. The first value represents the main decay time constant, while the second (in parentheses) corresponds to the long decay component. Bath and detector working temperatures were 60\,mK and 100\,mK, respectively.\\
  $^c$ Values reported are for the pixel with the best energy resolution. Bath and detector working temperatures were 60\,mK and 100\,mK, respectively.\\
  $^d$ $G$ was not measured for this array. Energy resolutions and time constants are for the best-performing pixel with minimal activity (about 0.04\,Bq). Bath and detector working temperatures were 40\,mK and 95\,mK, respectively.\\
  $^e$ $G$ was measured only for pixels in one half of the array. Energy resolutions and time constants are for the best-performing pixel with minimal activity (around 0.1\,Bq). Bath and detector working temperatures were 40\,mK and 95\,mK, respectively.\\
  \end{flushleft}
  \end{table*}

Pulses acquired at the optimal working point are triggered offline and then processed to reject spurious pulses, to correct for gain drifts, and to estimate the amplitude by optimal filtering \cite{gatti_processing_1986} and other parameters, such as pulse time constants \cite{borghesi_first_2022}. Pulse amplitude spectra are energy calibrated by interpolating the X-ray peaks with a quadratic model forced through the origin. The detector activity is estimated from the integral of the M1 peak\footnote{The M1 branching ratio is determined from a high-statistics measurement of the full spectrum. Detailed results on the full EC spectral shape will be discussed in a forthcoming publication \cite{ahrens_phenomenological_2025}.}.

Tab.\,\ref{tab:results} summarizes the properties and performance of the TES microcalorimeters in the three arrays discussed here and compares them with previously reported results \cite{alpert_high-resolution_2019b}.

\begin{figure}[bt!]
  \begin{center}
  \includegraphics[width=0.45\textwidth]{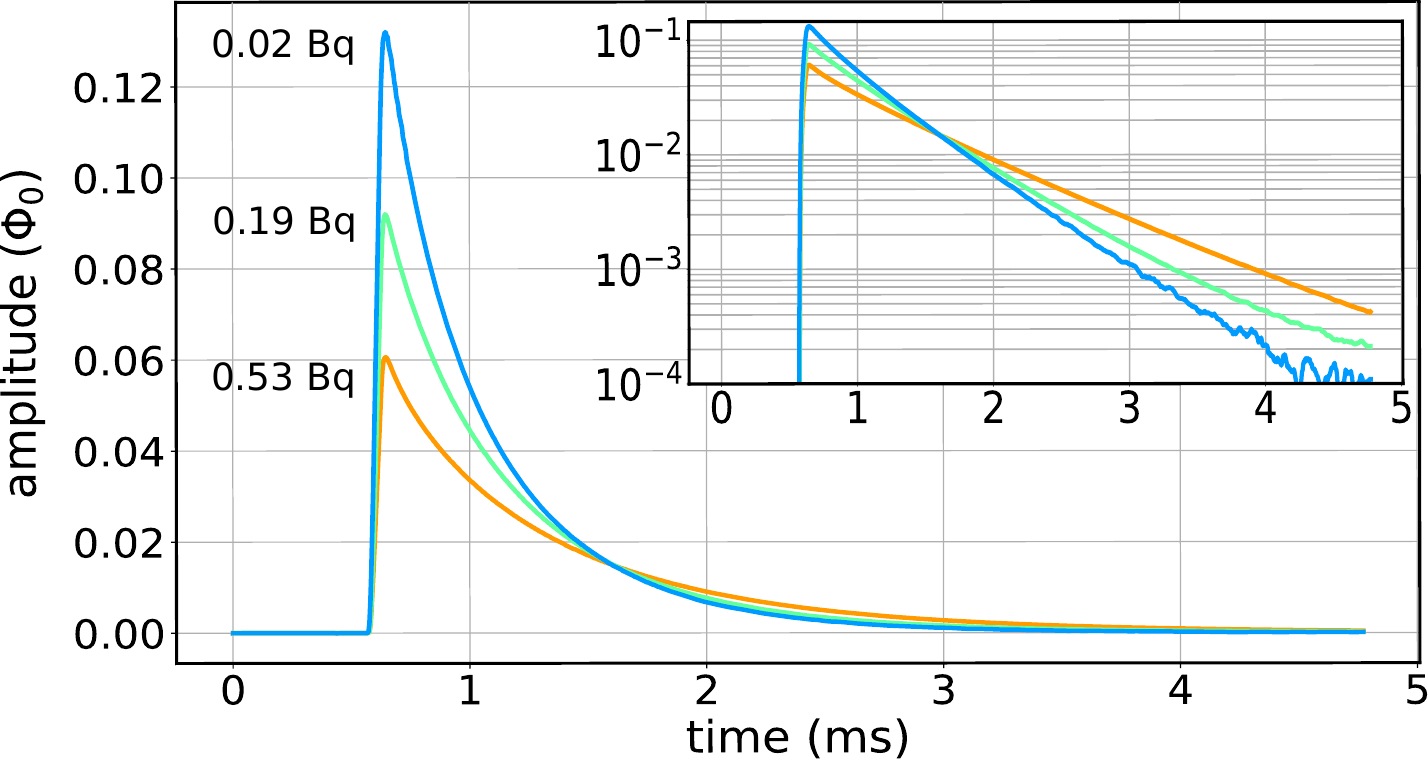}
  \end{center}
   \caption{Averaged pulses from N1 capture events in pixels with about 0.02\,Bq, 0.2\,Bq, and 0.5\,Bq of \Ho\ activity from Array 2. The logarithmic scale in the inset highlights the absence of additional long decay time constants. }
  \label{fig:pulses}
 \end{figure}

\section{Results}
 Pulses recorded at the N1 peaks (Fig.\,\ref{fig:pulses}) in the two implanted arrays are well described by a double exponential model with a rise and fall time. 
 However, a slight deviation from the simple exponential behavior is observed at the start of the decay, likely due to the breakdown of the small signal approximation. Furthermore, aside from the primary exponential decay time constants, which range between approximately 600\,\mus\ and 900\,\mus, no additional long time constants correlated with the implanted activities are observed in the pulses.

\begin{figure}[tb]
 \begin{center}
\includegraphics[width=0.4\textwidth]{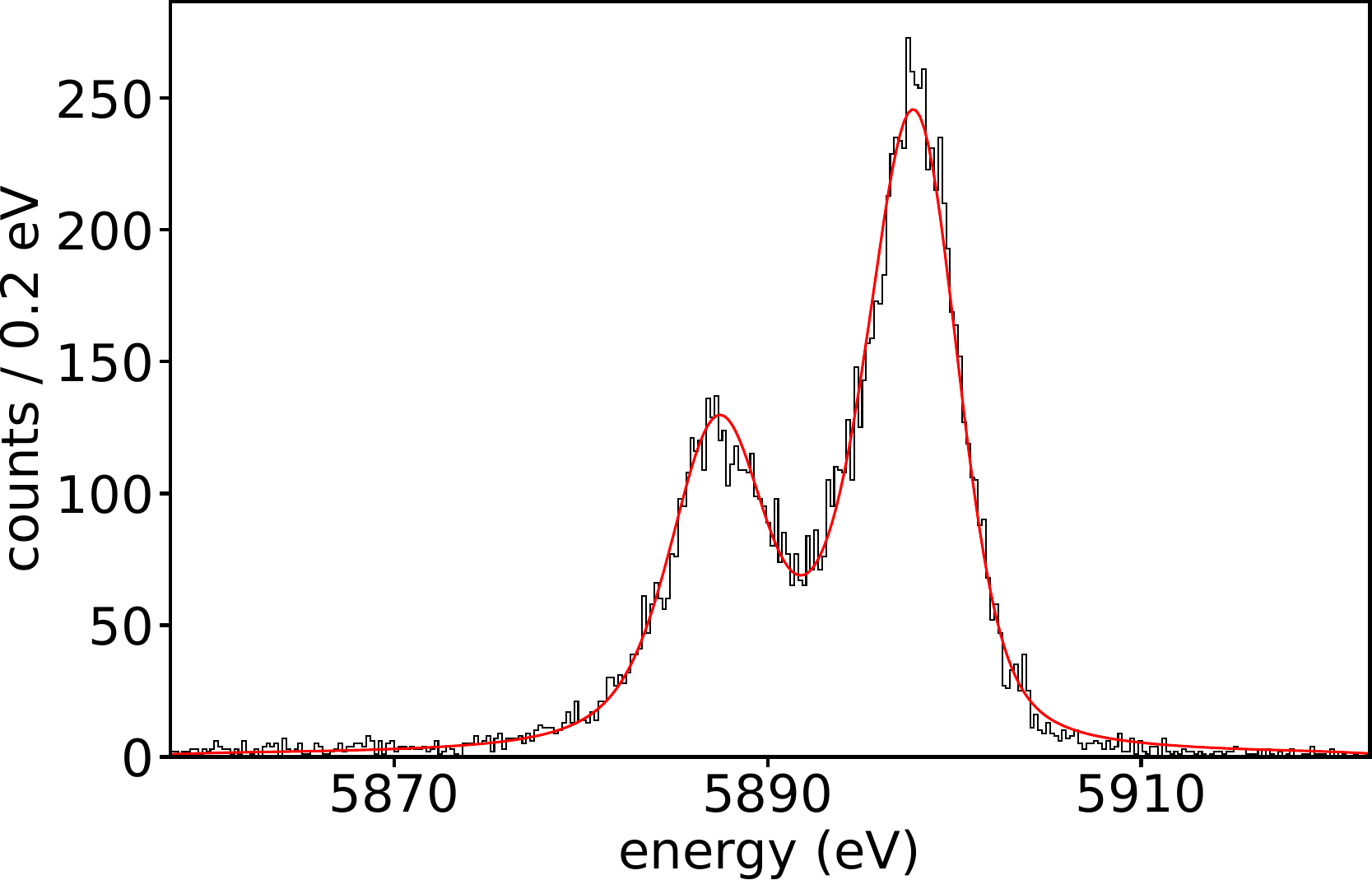}
\end{center}
 \caption{Mn K$\alpha$ X-ray peak measured with the best pixel in Array 0. The FWHM energy resolution is $(4.2\pm0.1)$\,eV, while the intrinsic resolution, calculated from Eq.\,(\ref{eq:nep}), is about 3.7\,eV.}
 \label{fig:spe0}
\end{figure}

\begin{figure}[bt]
 \begin{center}
 \includegraphics[width=0.48\textwidth]{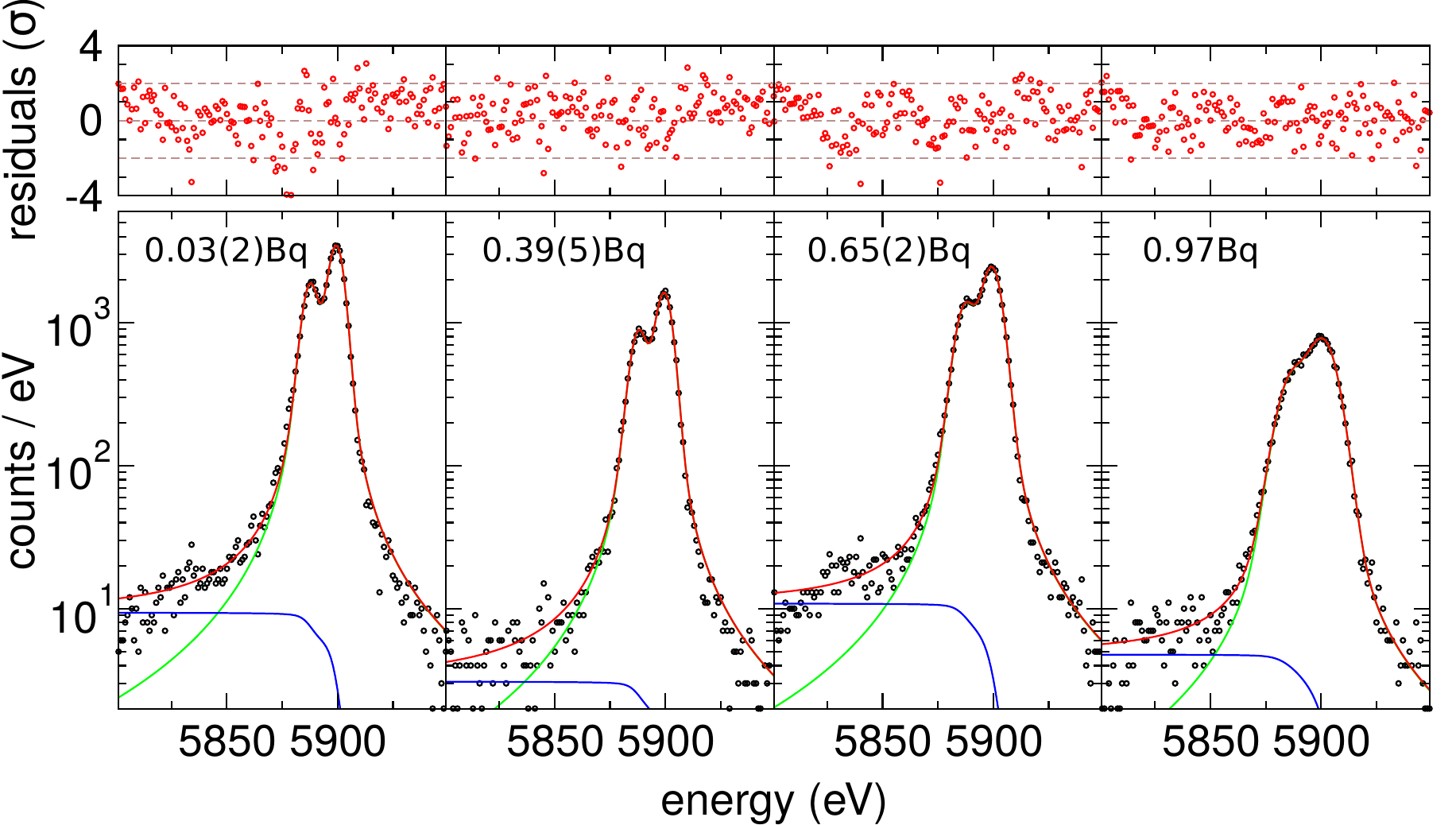}
 \end{center}
 \caption{The four panels from left to right show the Mn K$_\alpha$ peaks obtained summing spectra from a few pixels with increasing activity.
The FWHM energy resolutions obtained fitting the Mn K$_\alpha$ peaks from left to right are $6.5\pm0.1$, $6.9\pm0.1$, $8.5\pm0.1$ and $13.1\pm0.1$\,eV.
Red curves show the fitted model with Gaussian peaks for all seven Mn K$\alpha$ components \cite{holzer_$kensuremathalpha_12$_1997}; blue curves show the flat background; green curves represent the sum of all Gaussians.
 }
\label{fig:spe_a2}
\end{figure}

The energy resolution is evaluated using the Mn K$\alpha$ peak, leveraging its high-statistics data across all three arrays. For improved precision, the local energy scale is calibrated using the Mn K$\alpha_1$ and K$\alpha_2$ peaks \cite{holzer_$kensuremathalpha_12$_1997}. Fig.\,\ref{fig:spe0} shows the Mn K$\alpha$ peak resolution for the best pixel in Array 0.
Mn K$\alpha$ peak spectra from detectors with comparable activity and energy resolution were summed to improve statistics. The resulting spectra were then fitted using the method described in \cite{ferri_investigation_2012a} to investigate the presence of peak tails potentially associated with holmium implantation.
Refer to Fig.\,\ref{fig:spe_a2} for details on the spectra. The red curves represent the fitted model, including Gaussian peaks for all seven Mn K$\alpha$ components \cite{holzer_$kensuremathalpha_12$_1997}, while the blue curves show the flat background model, which correlates with the Mn K$\alpha$ peak intensity but not with the implanted activity. This background, caused by X-rays interacting outside the absorber, appears only on the low-energy side of the peaks and has a level of approximately $A_{tail}(\mathrm{K}\alpha)\times 0.006$\,counts/eV, where $A_{tail}$(K$\alpha$) is the total peak amplitude. The green curves show the sum of all Gaussian components. Including an exponential tail component as in \cite{ferri_investigation_2012a}, with a scale parameter $\lambda_{tail}$ ranging from 0.1\,eV to 100\,eV$^{-1}$, does not improve the fit. The tail amplitude is always negligible, at most 0.1\% of the total Gaussian amplitude, and shows no correlation with the implanted activity.

All TESs are found to be well in the extreme electro-thermal feedback (ETF) regime as
\begin{equation}
\mathcal{L}_I = \frac{I_0^2 R_0 \alpha_I}{G T_0} \gg 1
\end{equation}
where $\alpha_I$, the TES's logarithmic constant current temperature sensitivity, is not directly assessed for the TESs under test, but it is expected to be about 100 with a 30\% $1\sigma$ uncertainty.

The decay time constant is found fitting pulses from the N1 peak of the \Ho\ spectrum to remain in the small signal limit.
Considering also that the detectors are in the low inductance limit, the decay time constant is expected \cite{irwin_transition-edge_2005} to be
\begin{equation}
\label{eq:tau}
\tau_{dec} = \frac{C}{G} \frac{1+\beta_I+R_L/R_0}{1+\beta_I+R_L/R_0+(1-R_L/R_0)\mathcal{L}_I} = \eta \frac{C}{G}
\end{equation}
where $R_L$ and $\beta_I$ are the reference (or load) resistor and the TES's logarithmic current sensitivity at constant current.
Also $\beta_I$ is only known from the characterization of similar devices to be around 2 with a 50\% $1\sigma$ uncertainty.

In the strong ETF regime and with negligible amplifier noise, the detector intrinsic FWHM energy resolution is expected \cite{irwin_transition-edge_2005} to be
\begin{equation}
\label{eq:de}
\Delta E_\mathrm{0}=
2\sqrt{2 \mathrm{ln} 2}
\sqrt{4 k_\mathrm{B} T_{b}^2 \frac{C }{\alpha_I}\sqrt{n/2}}
\end{equation}
The intrinsic energy resolution for each pixel is calculated by integrating the power spectral density of the noise, $S_{pow}(f)$ \cite{irwin_transition-edge_2005}:
\begin{equation}
\label{eq:nep}
\Delta E_\mathrm{0} = 2 \sqrt{2 \ln 2} \sqrt{\int_0^{\infty} \frac{4}{S_{pow}(f)} \, \mathrm{d}f}.
\end{equation}
Here, $S_{pow}(f)$ is obtained by averaging the Fourier transform of noise waveforms and using the detector's current-to-power responsivity, derived from the average pulse shape and amplitude for known small energies (small signal limit).
In both eq.\,(\ref{eq:tau}) and eq.\,(\ref{eq:de}), $C$ is the total heat capacity of the detector at the working temperature $T_0$. To interpret the data we assume that
\begin{equation}
 \label{eq:c}
 C(A_\mathrm{Ho}) = C_\mathrm{TES} + C_\mathrm{Ho} = C_\mathrm{TES} + n_\mathrm{Ho} c_\mathrm{Ho} = C_\mathrm{TES} + \frac{A_\mathrm{Ho}}{\lambda N_A} c_\mathrm{Ho}
\end{equation}
where $C_\mathrm{TES}$ is the combined heat capacity of the TES sensor and the absorber -- estimated to be approximately $0.8\times10^{-12}$\,J/K at around 90\,mK and refined to $(1.0\pm0.3)$\,pJ/K based on Array 0 data using eq.\,(\ref{eq:tau}) -- and $n_\mathrm{Ho}$, $c_\mathrm{Ho}$, $A_\mathrm{Ho}$, and $\lambda$ represent the number of moles, the specific heat capacity in J/K/mol, the implanted activity in the detector in Bq, and the decay constant of \Ho\ ($4.8\times10^{-12}$\,s$^{-1}$), respectively.

\begin{figure}[!t]
 \begin{center}
 \includegraphics[width=0.430\textwidth]{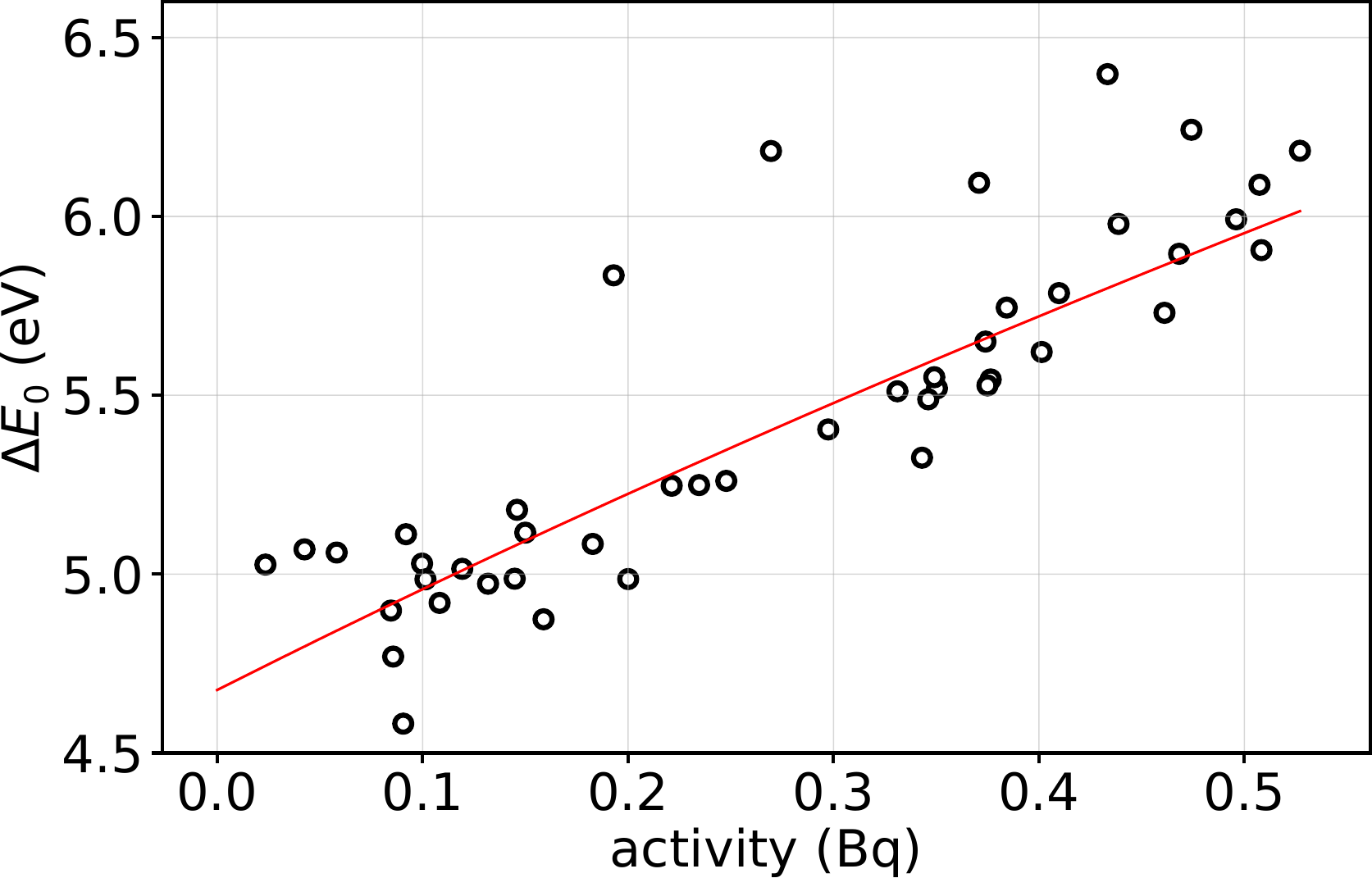}
 \end{center}
\caption{Plot of $\Delta E_\mathrm{0}$ versus $A_\mathrm{Ho}$ for Array 2. Intrinsic resolutions are derived using Eq.\,(\ref{eq:nep}) as described in the text. Pixel activities are calculated from the M1 peak integral, using its branching ratio from the full \Ho\ spectrum analysis, to be detailed in an upcoming publication. The solid line represents the robust fit to the data.}
  \label{fig:de_vs_a}
\end{figure}
Considering $C(A_\mathrm{Ho})$ as given by Eq.\,\ref{eq:c}, we extracted $c_\mathrm{Ho}$ by fitting $\Delta E_\mathrm{0}$ from Eq.\,(\ref{eq:nep}) as a function of $A_\mathrm{Ho}$ in Fig.\,\ref{fig:de_vs_a}, using the model $y(A) = a \sqrt{ C_\mathrm{TES} + b A_\mathrm{Ho}}$ (from Eq.\,\ref{eq:de}), where $a\propto \sqrt{T_{b}^2\sqrt{n}/\alpha_I}$ and $b=c_\mathrm{Ho}/(\lambda N_A)$ are free parameters. The solid line in Fig.\,\ref{fig:de_vs_a} shows the result of a robust fit, yielding $c_\mathrm{Ho} = \lambda N_A b = (2.9 \pm 0.4\,\mathrm{(stat)} \pm 0.7\,\mathrm{(sys)})$\,J/K/mol at $(94 \pm 1)$\,mK. This value is slightly lower than, but consistent with, the literature value for the heat capacity of metallic holmium (3.8\,J/K/mol \cite{krusius_calorimetric_1969}). 
The statistical uncertainty reflects pixel-to-pixel variations in the parameters of eq.\,(\ref{eq:de}), while the systematic uncertainty is dominated by the uncertainty in $C_\mathrm{TES}$, estimated to range between 0.6\,pJ/K and 1.0\,pJ/K.

\begin{figure} [!tb]
 \begin{center}
 \includegraphics[width=0.4\textwidth]{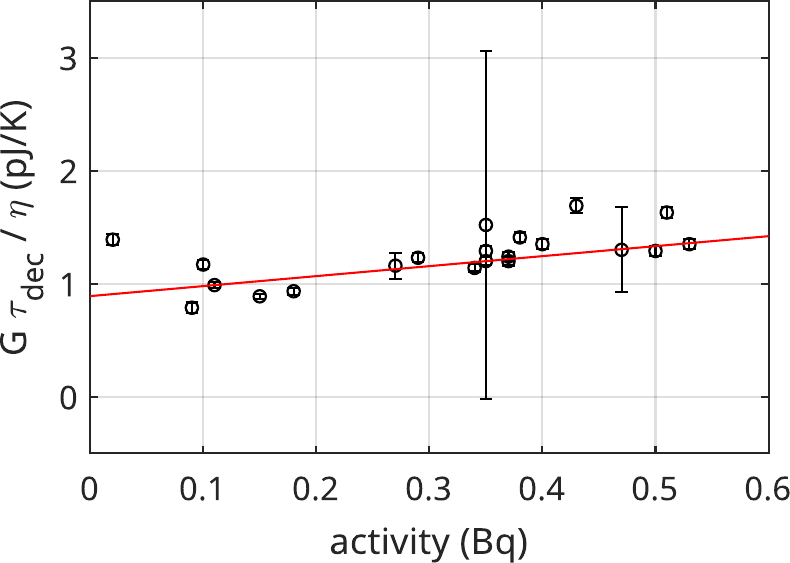}
 \end{center}
\caption{Plot of $\tau_{dec} G / \eta$ versus $A_\mathrm{Ho}$ for one half of the pixels in Array 2. Error bars reflect the propagated uncertainties of all parameters in eq.\,(\ref{eq:tau}). The solid line shows the result of a robust linear fit to the data.}
 \label{fig:g_vs_a}
\end{figure}
For the derivation of $c_\mathrm{Ho}$ from exponential decay time constant of N1 pulses, we plot $\tau_{dec} G / \eta$ versus the pixel activity $A_\mathrm{Ho}$ as in Fig.\,\ref{fig:g_vs_a}. Here the error bars are obtained propagating the errors on all parameters appearing in eq.\,(\ref{eq:tau}) and are dominated by the uncertainty on $\alpha_I$.
A robust linear interpolation gives $C_\mathrm{TES} = (0.89\pm0.08)$\,pJ/K and $c_\mathrm{Ho} = (2.5\pm 0.7)$\,J/K/mol.

\section{Discussion}
The values obtained for $c_\mathrm{Ho}$ are consistent with each other and align with those reported by ECHo \cite{herbst_specific_2021} and for metallic holmium \cite{krusius_calorimetric_1969}. The difference between the two estimates may arise from non-linear TES response and systematic shifts due to fixed $\alpha_I$ and $\beta_I$. Given its reduced sensitivity to these parameters, we consider the result derived from $\Delta E_0$ to be the most robust. 
Notably, the significance of our findings arises from the distinct physical mechanisms governing signal formation in TES detectors compared to previous studies \cite{herbst_specific_2021}.

\begin{figure}[bt]
  \begin{center}
  \includegraphics[width=0.48\textwidth]{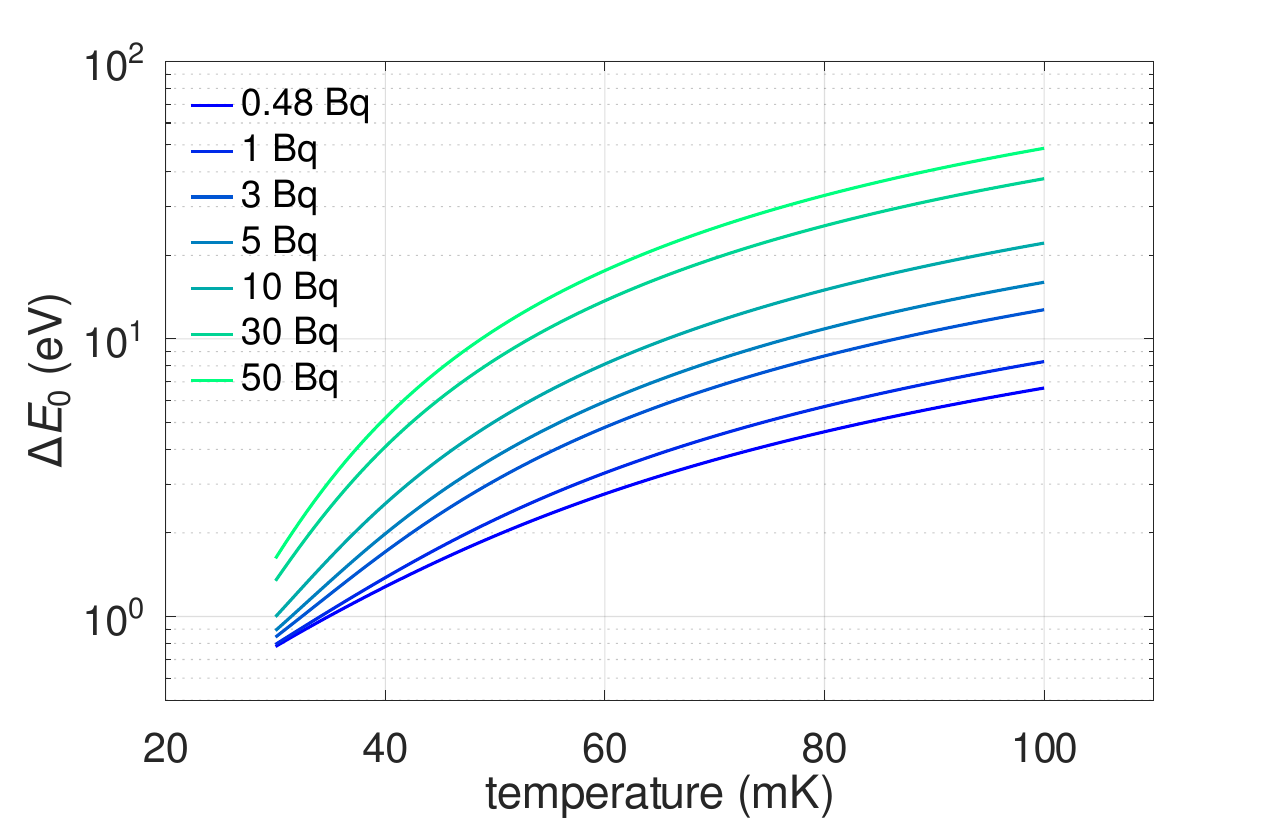}
  \end{center}
  \caption{Projected intrinsic energy resolution $\Delta E_\mathrm{0}$ as a function of TES operating temperature for different \Ho\ activities. Curves are normalized to 6\,eV FWHM at 0.48\,Bq and 90\,mK (average activity and energy resolution of Array 2 \cite{alpert_most_2025}). Calculations assume $C_\mathrm{TES}$ scales with $T$, $c_\mathrm{Ho}$ from \cite{krusius_calorimetric_1969}, and that detector design can be optimized to maintain $G$ and $\alpha_I$ at suitable values for each temperature and activity. In particular, $\alpha_I$ is assumed to remain at least 100 and the decay time constant is kept on the order of 100\,\mus.
}
  \label{fig:trend}
 \end{figure}

 Array 2 has been used for taking high statistics data with the aim of performing an analysis of the \Ho\ decay end point to derive a limit on the neutrino mass \cite{alpert_most_2025}. 
 Improving the neutrino mass sensitivity requires 
 measurements of even more decays through a combination of more detectors and higher per pixel \Ho\ activity, without significantly degrading energy resolution. 
  To achieve neutrino mass sensitivities on the order of 1\,eV$/c^2$, Monte Carlo simulations \cite{nucciotti_statistical_2014b} indicate that the impact on sensitivity remains negligible as long as the energy resolution is better than 10\,eV.
 The extrapolation of the fit in Fig.\,\ref{fig:de_vs_a} suggests that the energy resolution remains within this limit as long as the pixel activity is kept below a few decays per second. Further increases in activity require a reduction of the TES operating temperature.
Fig.\,\ref{fig:trend} illustrates the expected scaling of the intrinsic energy resolution $\Delta E_\mathrm{0}$ with TES operating temperature and implanted $^{163}$Ho activity, based on the assumptions detailed in the caption.
Targeting sub-eV neutrino mass sensitivities will require large arrays with pixel activities of tens or even hundreds of decays per second, while maintaining an energy resolution as close as possible to 1\,eV. From Fig.\,\ref{fig:trend}, it is apparent that operating detectors with about 50\,Bq at the required performance level demands reducing the operating temperature below 40\,mK. In this context, based on current knowledge, managing activities significantly above 50\,Bq -- as originally proposed in \cite{alpert_holmes_2015b} -- appears challenging.

\section{Conclusions}
In summary, this work demonstrates that embedding \Ho\ in the absorber of TES microcalorimeters at activities up to approximately 1\,Bq results solely in an increased heat capacity, with no evidence of non-Gaussian response or additional thermally decoupled systems. Despite the relatively low \Ho\ concentration in the gold absorber, the measured specific heat capacity is consistent with values reported for metallic holmium. Extrapolation of our results indicates that, to achieve detector activities as high as 50\,Bq without significant degradation in performance, the microcalorimeters must be redesigned to operate at temperatures as low as 40\,mK.

\begin{acknowledgements}
This work and the HOLMES experiment have been supported by Istituto Nazionale di Fisica Nucleare (INFN) and by European Research Council under the European Union’s Seventh Framework Programme (FP7/2007-2013)/ERC Grant Agreement no. 340321.
\end{acknowledgements}

\section*{Declarations}

\textbf{Conflict of interest} The authors have no competing interests to declare that are relevant to the content of this article.\\ \\
\textbf{Availability of data and materials} Data available from the corresponding authors upon reasonable request.\\ \\
\textbf{Authors' contributions}\\
Conceptualization: Matteo Borghesi, Matteo De Gerone, Marco Faverzani, Elena Ferri, Flavio Gatti, Andrea Giachero, Lorenzo Ferrari Barusso, Angelo Nucciotti, Luca Origo, Dan Schmidt, Joel Ullom, Sara Gamba;
Investigation: Matteo Borghesi, Marco Faverzani, Elena Ferri, Sara Gamba, Luca Origo;
Formal Analysis: Matteo Borghesi, Elena Ferri, Luca Origo, Angelo Nucciotti;
Software: Matteo Borghesi, Pietro Campana, Rodolfo Carobene, Joseph Fowler, Sara Gamba, Andrea Giachero, Roberto Moretti, Luca Origo;
Data Curation: Matteo Borghesi, Marco Faverzani, Elena Ferri;
Visualization: Marco Gobbo, Danilo Labranca, Luca Origo;
Resources: Douglas Bennett, Giancarlo Ceruti, Matteo De Gerone, Lorenzo Ferrari Barusso, Dan Schmidt;
Supervision: Flavio Gatti, Stefano Ragazzi, Daniel Swetz, Joel Ullom, Angelo Nucciotti;
Funding Acquisition: Matteo De Gerone, Marco Faverzani, Flavio Gatti, Stefano Ragazzi, Joel Ullom;
Project Administration: Angelo Nucciotti;
Writing-Original Draft Preparation: Marco Faverzani, Elena Ferri, Andrea Giachero, Angelo Nucciotti;
Writing-Review \& Editing: Douglas Bennett, Matteo Borghesi, Pietro Campana, Rodolfo Carobene, Matteo De Gerone, Joseph Fowler, Sara Gamba, Marco Gobbo, Danilo Labranca, Roberto Moretti, Angelo Nucciotti, Dan Schmidt, Daniel Swetz, Joel Ullom;

\appendix
\section{\Ho\ isotope embedding}
\label{app:implant}
The \Ho\ nuclei for the HOLMES experiments are produced by irradiating \Er\ enriched \ero\ samples in a flux of thermal neutrons \cite{heinitz_production_2018b}. The irradiated samples are then chemically purified by ion exchange chromatography to separate other chemical species produced by nuclear reactions in the reactor \cite{heinitz_production_2018b}. The final samples still contain 
traces of $^{165}$Ho ($\approx 2$ $^{163/165}$Ho at. ratio) and $^{166m}$Ho ($\approx 2\times10^3$ $^{163/166m}$Ho at. ratio) which must be removed to prevent extra contributions to the detector heat capacity and, in the case of the beta decaying $^{166m}$Ho, to limit background counts interfering with the neutrino mass measurement.
The \Ho\ nuclei are introduced in HOLMES microcalorimeters by ion implantation \cite{degerone_development_2023}. A holmium ion current is produced by a hot-running cold plasma sputter ion source. Ions are accelerated to 30\,keV and the 163 mass component of the beam is selected by a solenoid magnet and an adjustable slit at the beam waist. When it reaches the detector array, the beam has an approximately 2D gaussian cross-section with a FWHM of about 4\,mm.  The deflected 165 and 166 mass ions are separated from the 163 mass one by about 7 and 11 $\sigma$s, respectively, and are almost completely masked by the slit.
Based on SRIM\footnote{SRIM 2013 software package, \url{https://www.srim.org}} simulations of the implantation profile -- which has a mean depth of about 4\,nm for 30\,keV ion energy -- we estimate an approximate average (peak) holmium concentration of 
$x_{\mathrm{Ho}} \approx 1.5\%(1.6\%)$ for a \Ho\ activity of 1\,Bq implanted in a $180\times180$\,\mum$^2$ gold absorber.

Arrays 1 and 2 (see Section\,\ref{sec:exp_meth}) differ in their ion implantation protocols. Array 1 was ion-implanted with a single central shot, targeting a nominal peak activity of approximately 4\,Bq. Array 2 received four implantation shots, designed to achieve a more uniform activity distribution with a nominal maximum of about 2\,Bq. 
Due to the unavoidable sputtering effect of the ion beam, SRIM simulations indicate that the actual detector activity should saturate at around 2\,Bq for a $180\times180$\,\mum$^2$ gold absorber and an ion beam energy of 30\,keV.
However, for reasons not yet fully understood, the highest measured activity was approximately 1\,Bq.
Each beam shot was performed with \Ho\ ion currents of approximately 10\,nA for several thousand seconds.

\section{Microcalorimeter array fabrication}
\label{app:fab}
HOLMES microcalorimeters are arranged in 64 pixel arrays fabricated at different institutions of the HOLMES collaboration \cite{borghesi_first_2022}. 
NIST carries out the fabrication without performing the silicon micromachining to release the SiN membranes with the detectors on top. At this stage, only the bottom gold layer of the absorber, 1\,\mum\ thick, is deposited.
The wafer is then cut into $14\times19$\,mm$^2$ chips containing two arrays each. To allow further processing the chips have  masks on both sides.
On the detector side, a photoresist lift-off mask is left to complete the detector absorbers. On the back of the chip a SiN mask is patterned to allow anisotropic silicon etching to release the membranes.
Subsequent fabrication steps are performed at the INFN laboratories of the Universities of Genova and Milano-Bicocca.
After ion implanting the \Ho\ in the bottom gold layer with a depth profile extending about 10\,nm, the top gold layer -- 1\,\mum\ thick -- is deposited in an ion beam assisted sputtering system. The use of four symmetric ion beams and gold targets gives a uniform -- 4\% spread on a $10\times 10$\,mm$^2$ sample -- gold deposition on the array surfaces at a rate of more than 100\,nm/h. After the second Au layer is completed and the \Ho\ is completely encapsulated, the chip is dipped in an acetone bath at 50\,°C for 2\,hours to perform the lift-off process. Finally, the chip is mounted in a EPDM o-ring sealed POM holder with only the back exposed and it is dipped in a KOH bath at 80\,°C for about 5\,hours, until the SiN membranes are completely released by the anisotropic etching.
Although the current fabrication process of the absorber does not provide full lateral containment of the radiation emitted by the implanted source (Fig.\,\ref{fig:tes}), Monte Carlo simulations show that this has a negligible effect on the measured activity.
As shown in Table\,\ref{tab:results}, and corroborated by the decay time constants, all three arrays exhibit a $G$ smaller than expected based on the data reported in \cite{alpert_high-resolution_2019b}, with significant variation across the arrays. This discrepancy is likely attributable to the KOH processing used for these devices, as opposed to the DRIE process employed in \cite{alpert_high-resolution_2019b}. While the exact mechanism remains unclear, the variation appears to stem from bubble formation and trapping during the silicon etching in the KOH soaking process.

\section{Array readout}
\label{app:readout}
These arrays were mounted in gold-plated copper boxes (see Fig.\,\ref{fig:holders}), which also house the circuitry for biasing the detectors, along with the multiplexing chips for signal readout. The electrical connections between the multiplexing chips are established using superconducting aluminum bonding wires, each with a diameter of 25 $\mu$m. To ensure proper thermal contact between the sub-array and the copper holder, $\sim$20 gold bonding wires, each 50 $\mu$m in diameter, were also employed.

The signals from the voltage-biased TES arrays are frequency-multiplexed in the 4-8\,GHz band using flux ramp-linearized rfSQUIDs \cite{becker_working_2019b}. 
Each multiplexing chip handles 32 detector channels, distributing them across a 512\,MHz bandwidth.
Signal demodulation of the 32 signals leverages a custom IF board for up- and down-conversion combined with a Software Defined Radio (SDR) implemented in the FPGA of a Reconfigurable Open Architecture Computing Hardware (ROACH2) board \cite{mchugh_readout_2012}  equipped with ADC/DAC
modules.
Probe tones are amplified at 4\,K by a LNA HEMT amplifier.

\bibliographystyle{spphys}
\bibliography{holmes-TES,holmes-bibliography}

\end{document}